\documentclass[aps,prl,reprint,showpacs,preprintnumbers,amsmath,amssymb,superscriptaddress]{revtex4-1}

\usepackage{graphicx}
\usepackage{subcaption}

\begin{document} 

\frenchspacing

\title{Colossal pressure-induced softening in scandium fluoride}

\author{Zhongsheng Wei}
\affiliation{School of Physics and Astronomy, Queen Mary University of London, Mile End Road, London, E1 4NS, UK}

\author{Lei Tan}
\affiliation{School of Physics and Astronomy, Queen Mary University of London, Mile End Road, London, E1 4NS, UK}

\author{Guanqun Cai}
\affiliation{School of Physics and Astronomy, Queen Mary University of London, Mile End Road, London, E1 4NS, UK}

\author{Anthony E Phillips}
\affiliation{School of Physics and Astronomy, Queen Mary University of London, Mile End Road, London, E1 4NS, UK}

\author{Ivan da Silva}
\affiliation{ISIS Neutron and Muon Facility,  Rutherford Appleton Laboratory, Harwell Campus, Didcot, Oxfordshire, OX11 0QX, United Kingdom}

\author{Mark G Kibble}
\affiliation{ISIS Neutron and Muon Facility,  Rutherford Appleton Laboratory, Harwell Campus, Didcot, Oxfordshire, OX11 0QX, United Kingdom}

\author{Martin T Dove}
\email[Corresponding author: ]{martin.dove@qmul.ac.uk}
\affiliation{College of Computer Science, Sichuan University, Chengdu, Sichuan 610065, People's Republic of China}
\altaffiliation{Department of Physics, School of Sciences, Wuhan University of Technology, 205 Luoshi Road, Hongshan district, Wuhan, Hubei, 430070, People's Republic of China}

\begin{abstract}
The counter-intuitive phenomenon of pressure-induced softening in materials is likely to be caused by the same dynamical behaviour that produces negative thermal expansion. Through a combination of molecular dynamics simulation on an idealised model and neutron diffraction at variable temperature and pressure, we show the existence of extraordinary and unprecedented pressure-induced softening in the negative thermal expansion material scandium fluoride, ScF$_3$, with values of the pressure-derivative of the bulk modulus $B$, $B^\prime = \partial B / \partial P$, reaching as low as $-40 \pm 1$.
\end{abstract}

\maketitle

% In setting up this template for *Science* papers, we've used both
% the \section* command and the \paragraph* command for topical
% divisions.  Which you use will of course depend on the type of paper
% you're writing.  Review Articles tend to have displayed headings, for
% which \section* is more appropriate; Research Articles, when they have
% formal topical divisions at all, tend to signal them with bold text
% that runs into the paragraph, for which \paragraph* is the right
% choice.  Either way, use the asterisk (*) modifier, as shown, to

%\section*{Introduction}

Just over twenty years ago a seminal paper on negative thermal expansion in ZrW$_2$O$_8$ \cite{Mary:1996uq} marked the beginning of a new area in materials chemistry and physics that has led to the discovery of many materials with highly-anomalous physical properties. Since then we have seen not only have we seen an expansion in the range of materials showing negative thermal expansion, but also increases in the the size of expansivities  \cite{Takenaka:2012cv,Romao:2013ch,Dove:2016bv}. For example, around ten years ago, record positive \textit{and} negative expansivities were reported in the network material Ag$_3$Co(CN)$_6$ \cite{Goodwin:2008gb}. Other counter-intuitive properties related to atomic structure include negative linear compressibility \cite{Cairns:2015iw} and negative Poisson's ratio \cite{Alderson:2007bg}. More recently we have identified a new negative structural property, namely \textit{pressure-induced softening}, in which the material becomes softer (less stiff) under pressure \cite{Fang:2013ji,Fang:2013fj,Fang:2014cp,Fang:2013gp}. In this paper we report a new instance of this property with an extremely large, indeed unprecedented, negative size of the associated coefficient.

\begin{figure*}[tb]
\begin{center}
\begin{subfigure}[b]{0.34\textwidth}
\includegraphics[width=0.95\textwidth]{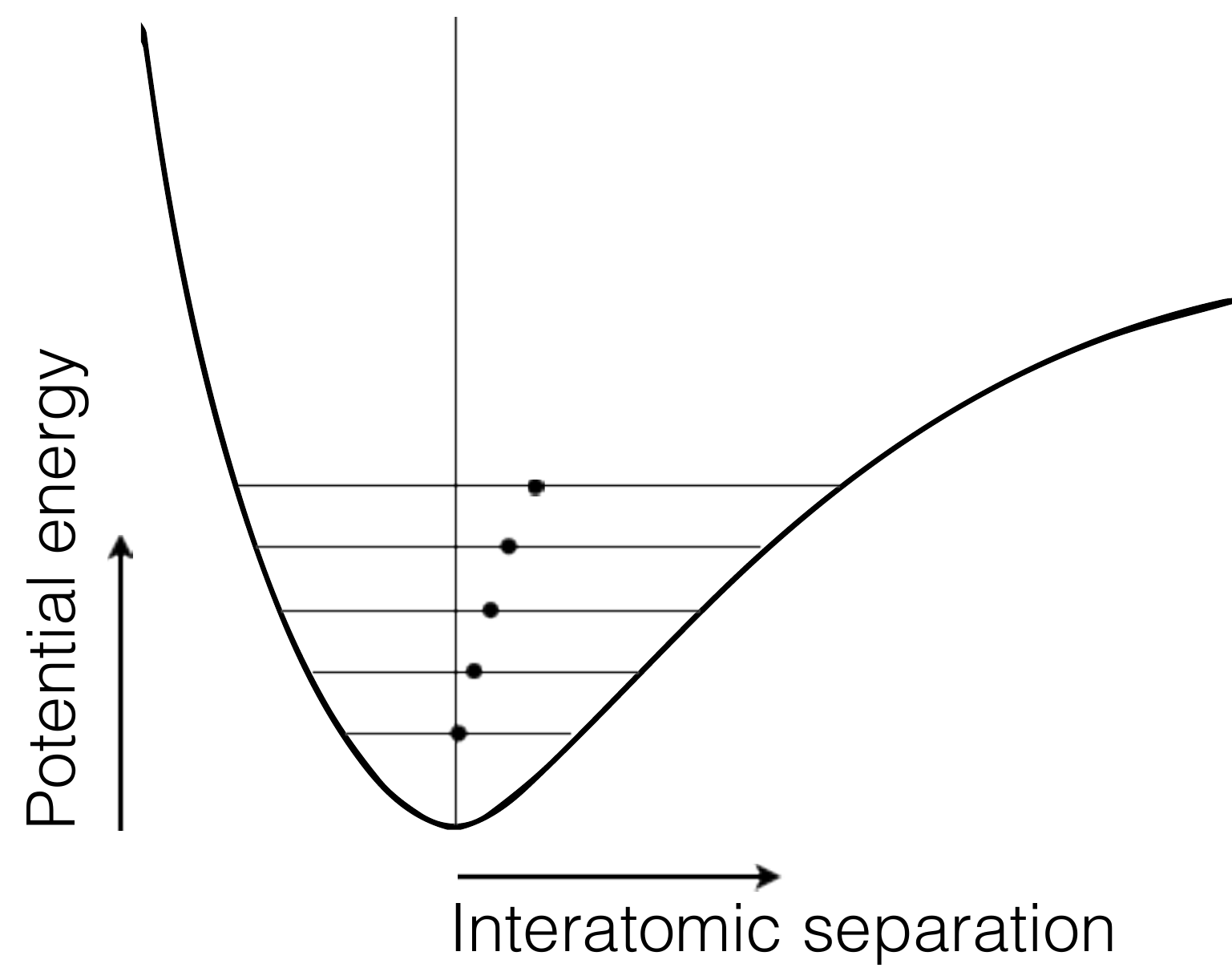}
\subcaption{}
\label{fig:bond_potential}
\end{subfigure}
\begin{subfigure}[b]{0.34\textwidth}
\includegraphics[width=0.8\textwidth]{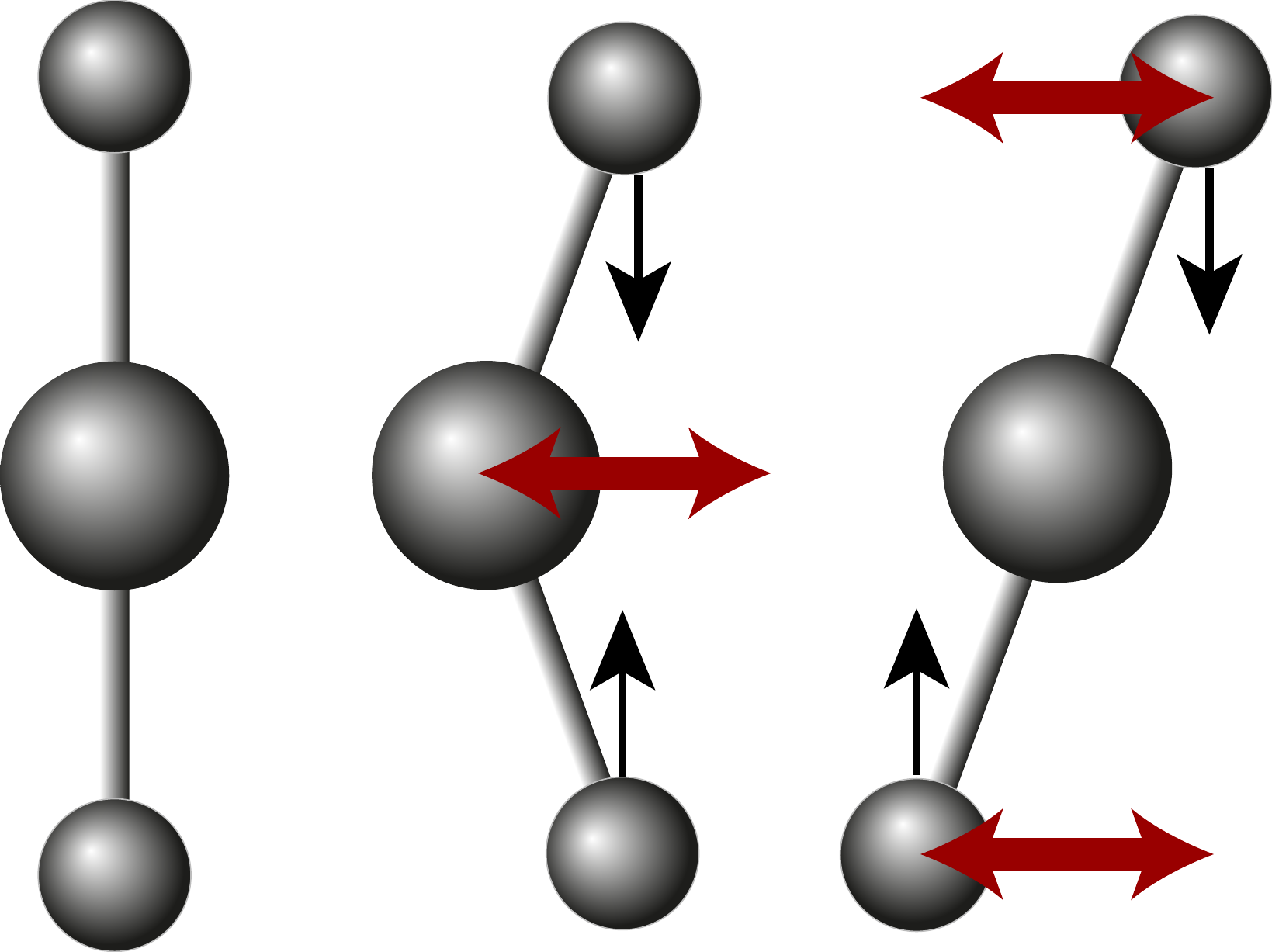}
\subcaption{}
\label{fig:tension_effect}
\end{subfigure} 
\begin{subfigure}[b]{0.24\textwidth}
\includegraphics[width=0.95\textwidth]{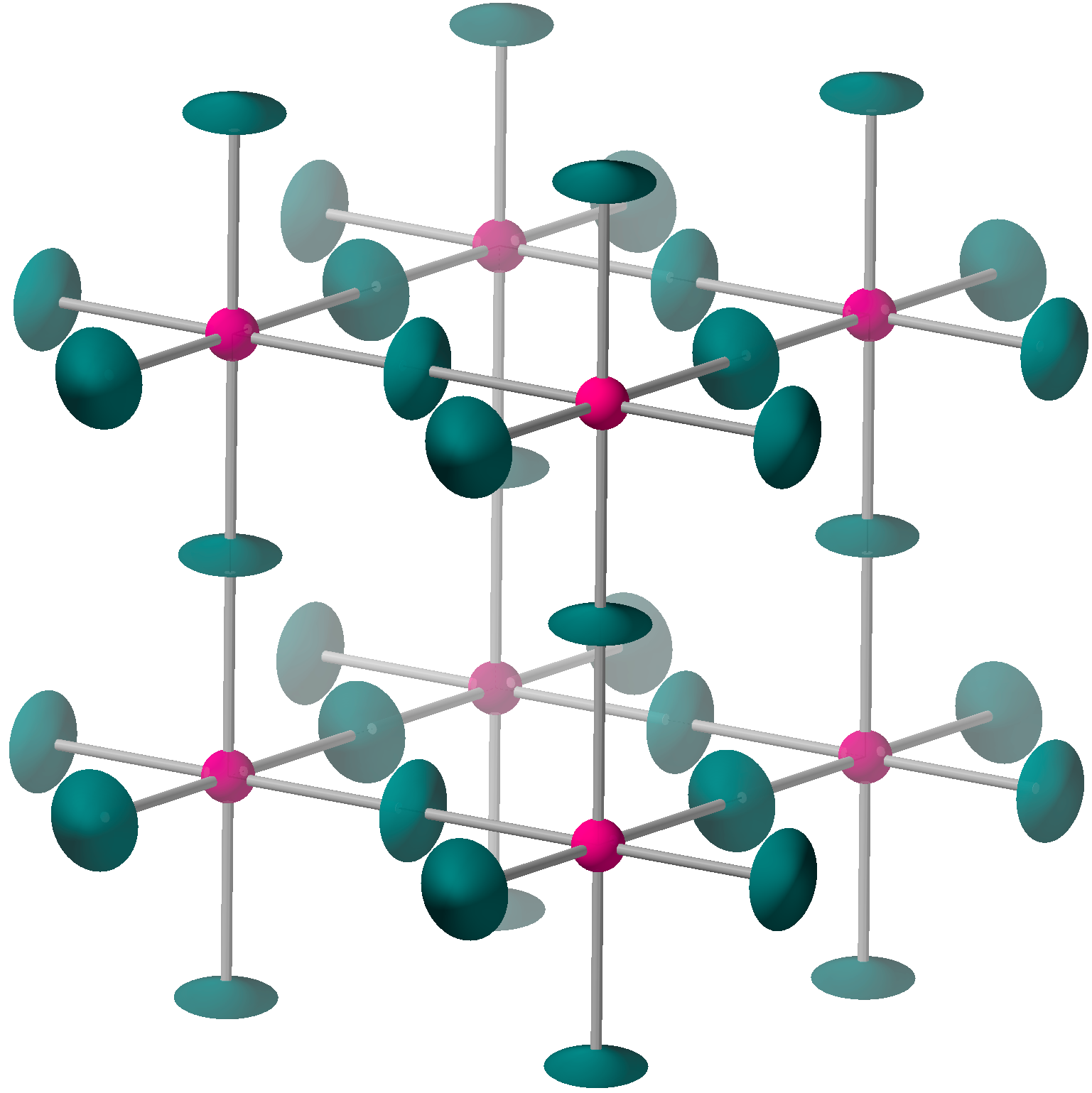}
\subcaption{}
\label{fig:crystal_structure}
\end{subfigure}
\caption{a) Representation of the potential energy of a bond between two atoms as a function of separation $r$. The diagram indicates the spread of bond lengths at higher temperatures, showing that the mean separation increases with temperature. b) Illustration of the tension effect, showing how transverse motions of an atom connected to two other atoms will pull the neighbouring atoms together if the motions do not stretch the bonds. c) Crystal structure of ScF$_3$ showing transverse motions of the F atoms by representing the atoms as ellipsoids whose major axes reflect the thermal motion.}
\end{center}
\end{figure*}

Negative thermal expansion is counter-intuitive because of the way we understand the forces between atoms. A chemical bond always (to the best of our knowledge) shows positive thermal expansion because it is harder to push two atoms together from the equilibrium distance than it is to pull them apart. Experimentally this has been observed in a number of materials, including the title material ScF$_3$ \cite{SiObond, SrTiO3_RMC, SrSnO3_RMC, Ag3CoCN6_RMC, Hu:2016it, Wendt:2019it}. The asymmetry in the distance-dependence of the bond potential means that at high temperature the vibration of the bond involves the bond stretching more than compressing, leading to an increase in the average separation, as illustrated in Figure \ref{fig:bond_potential}. This positive thermal expansion of the chemical bond usually leads to positive thermal expansion of the whole crystal. One common reason for negative thermal expansion of a crystal is called the ``tension effect', which is illustrated in Figure \ref{fig:tension_effect}. A transverse motion of an atom strongly-bonded to two neighbours on opposite sides will pull their mean positions inward, thereby reducing the average separation. This effect increases with temperature, leading to a negative expansivity, even with positive expansivities of the bonds. For a material such as the subject of this paper, ScF$_3$, the transverse atomic motions associated with the tension effect are seen in the atomic displacement parameters, as shown in Figure \ref{fig:crystal_structure}. We recently quantified this using neutron scattering methods \cite{Dove:2019tm}.

Pressure-induced softening is also counter-intuitive, and for similar reasons. The asymmetry of the potential energy of the chemical bond means that the more a bond is compressed, the harder it is to compress further, and the more the bond is stretched the easier it is to stretch it further. This means that we expect any crystal to become stiffer as it is compressed. An example from our human life is the snowball; it is soft when first formed, but then compression in the hand turns it into a hard sphere. However, in 1998 it was observed that amorphous silica shows the opposite behaviour over the pressure range 0--1.5~GPa \cite{Tsiok:1998hf}. It is as if the snowball gets softer the more we squash it! The resistance to compression shown by a material is quantified starting with the bulk modulus, defined as $B = - V \partial P/\partial V$, and its pressure dependence, $B^\prime = \partial B / \partial P$, is our quantity of interest here. A negative value of $B^\prime$ indicates pressure-induced softening \cite{Fang:2013ji,Fang:2013fj,Fang:2014cp,Fang:2013gp}. We previously reproduced the experimental negative value of $B^\prime$ in amorphous silica \cite{Tsiok:1998hf} using molecular dynamics simulations \cite{Walker:2007fp}, and showed that the effect is strongly correlated with the same sort of structural fluctuations that give rise to negative thermal expansion. 

\begin{figure}[tb]
\includegraphics[width=0.5\textwidth]{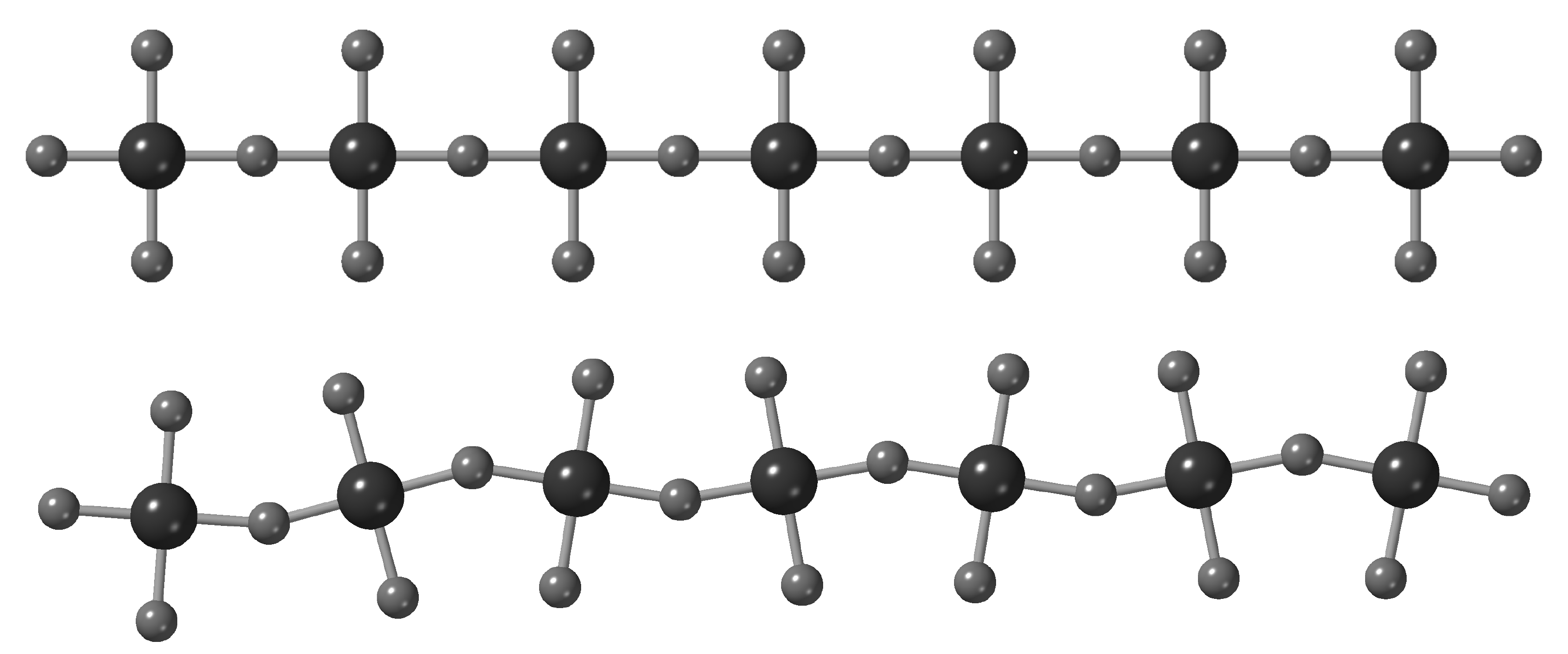}
\caption{The top shows a perfect alignment of atomic octahedra as in ScF$_3$. In this arrangement compression along the horizontal axis involves compression of the bonds, and likewise stretching of the arrangements involves stretching of the bonds. The bottom arrangement shows an elevated-temperature configuration where the thermal motion has rotated the octahedra. Compression or stretching now involve rotations of the bonds. }
\label{fig:cartoon}
\end{figure}

A simple explanation for the origin of pressure-induced softening is presented in Figure \ref{fig:cartoon}, which shows two linear arrangements of octahedra. In the case of perfect alignment, compression or stretching of the system involves compressing or stretching the bonds between atoms. If these bonds are strong -- which they usually are -- the elastic modulus is large. At a raised temperature, thermal atomic motion involves crumpling of the the arrangement. In this case, compression is made easier because it involves a mixture of further rotations of the bonds and compression. In general the force required to change the bond angle between octahedra through rotation is much lower than the the force required to compress bonds. This point has been discussed elsewhere in the context of the Rigid Unit Mode model \cite{Giddy:1993ue,Hammonds:1996wy}, a set of ideas related to how the flexibility of the network of corner-shared structural polyhedra (such as SiO$_4$ tetrahedra in silicates, TiO$_6$ octahedra in perovskites, and ScF$_6$ octahedra in the title material) impacts on the nature of the lattice vibrations and thence on behaviour such as phase transitions \cite{Hammonds:1996wy, Tridymite_RUMs} and negative thermal expansion \cite{Heine:1999vk}. With large rotations, the projection of the longitudinal compressional force resolves more onto causing octahedral rotations than onto compression of individual bonds. Larger external pressure then causes even larger rotations, and the material becomes even softer. This is pressure-induced softening. Similarly, we can consider stretching the crumpled chain: expansion first reduces the degree of rotation, until it reaches the point where further stretching of the system can only be accommodated by stretching the bonds, so that the system becomes elastically stiffer on stretching. This opposite perspective of becoming stiffer on reducing pressure is of course the same thing, but in our mind it is perhaps easier to visualise.

Pressure-induced softening has been observed in two materials that show negative thermal expansion, ZrW$_2$O$_8$ \cite{Drymiotis:2004kc} and Zn(CN)$_2$ \cite{Chapman:2007fg}. We were able to simulate pressure-induced softening in Zn(CN)$_2$ by the molecular dynamics method \cite{Fang:2013ji}, reproducing the experimental result. We also predicted a particular temperature dependence of $B^\prime$, which we subsequently confirmed experimentally \cite{Fang:2013fj}. Using an analytical model we showed that pressure-induced softening may be \textit{inevitable} -- rather than merely \textit{possible} -- in materials that show negative thermal expansion \cite{Fang:2014cp}, and in a simulation of several zeolites we showed that this prediction is reasonable \cite{Fang:2013gp}.

One material that shows negative thermal expansion and which has attracted a lot of interest is ScF$_3$ \cite{Greve:2010bu, Li:2011dn, Morelock:2013gi,Handunkanda:2015dc, Wang:2015gm, Wang:2015eo, Hu:2016it, Oba:2019bi, Dove:2019tm}. The crystal structure is shown in Figure \ref{fig:crystal_structure}; it is so irreducibly simple that it is effectively one of the few possible three-dimensional generalisations of the standard one-dimensional diatomic chain.\footnote{The other being the simple rocksalt structure.} We developed a very simple atomistic model for the dynamics of the ScF$_3$ structure, with a single bond-stretching interaction and two bond-bending interactions; details are given in the Supplementary Materials. Parameters were tuned by comparing the predictions with the results of ab initio calculations \cite{Li:2011dn}. Molecular dynamics simulations of this model showed negative thermal expansion as expected (Figure S1).

\begin{figure}[t]
\begin{center}
\begin{subfigure}[b]{0.48\textwidth}
\includegraphics[height=5cm]{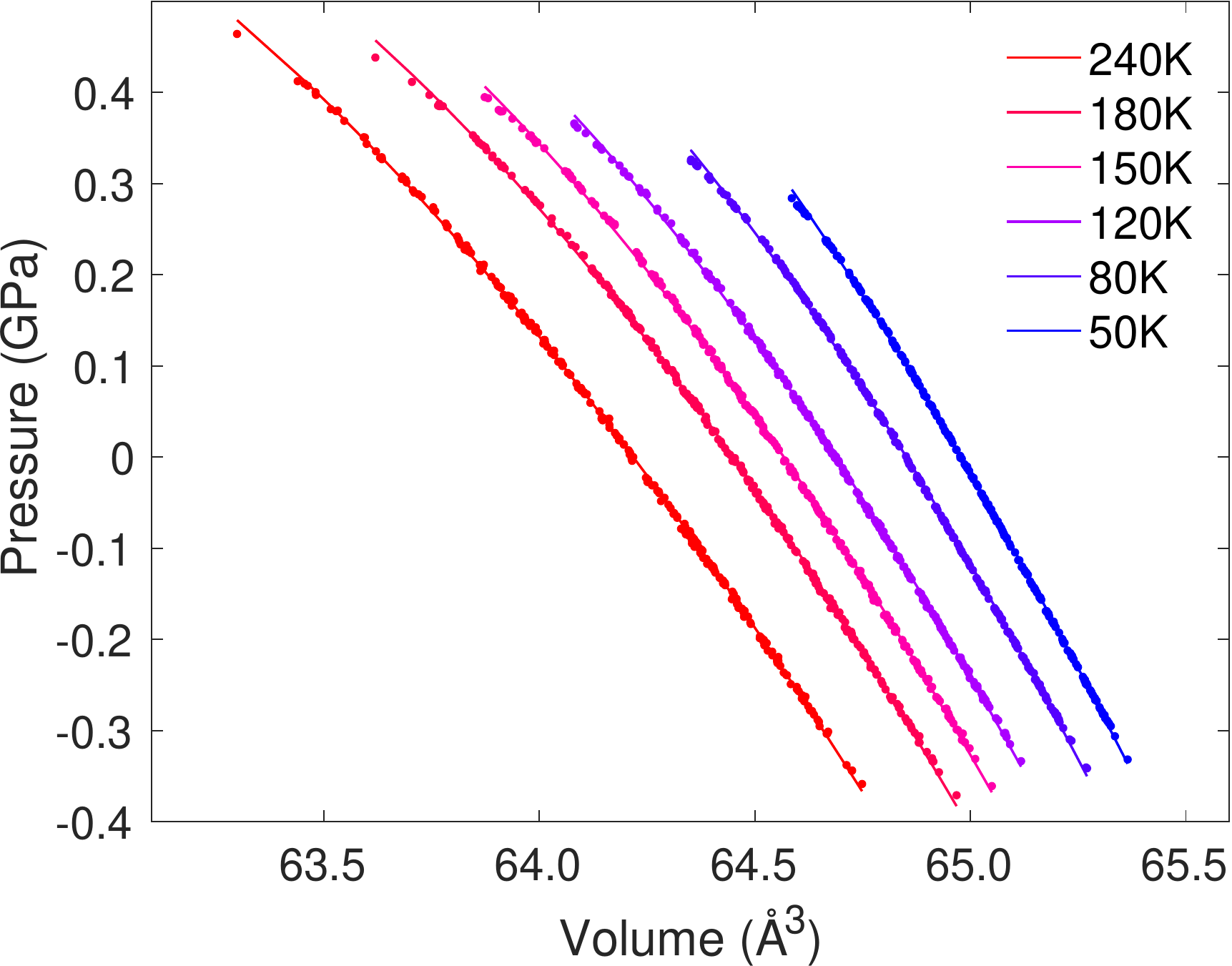}
\subcaption{}
\label{fig:PV_MD}
\end{subfigure}
\begin{subfigure}[b]{0.48\textwidth}
\includegraphics[height=5.05cm]{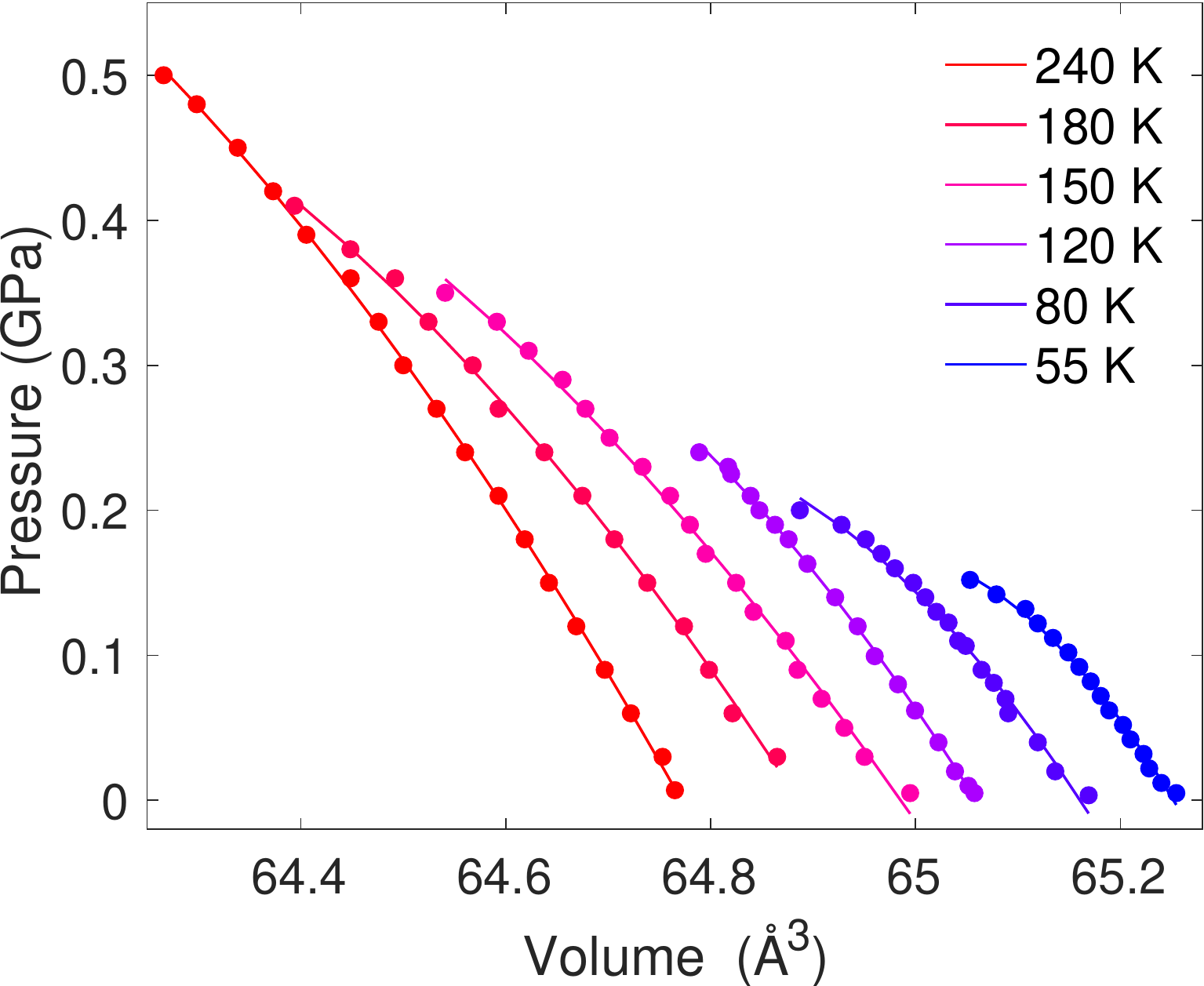}
\subcaption{}
\label{fig:PV_diffraction}
\end{subfigure}
\caption{$P(V)$ curves showing data from molecular dynamics simulation (a) and diffraction experiment (b) shown as filled circles, with the fitted third-order Birch-Murnaghan equation of state (equation \ref{eq:BMeos}) shown as curves.}
\end{center}
\end{figure}

\begin{figure}[t]
\begin{center}
\begin{subfigure}[b]{0.48\textwidth}
\includegraphics[width=0.9\textwidth]{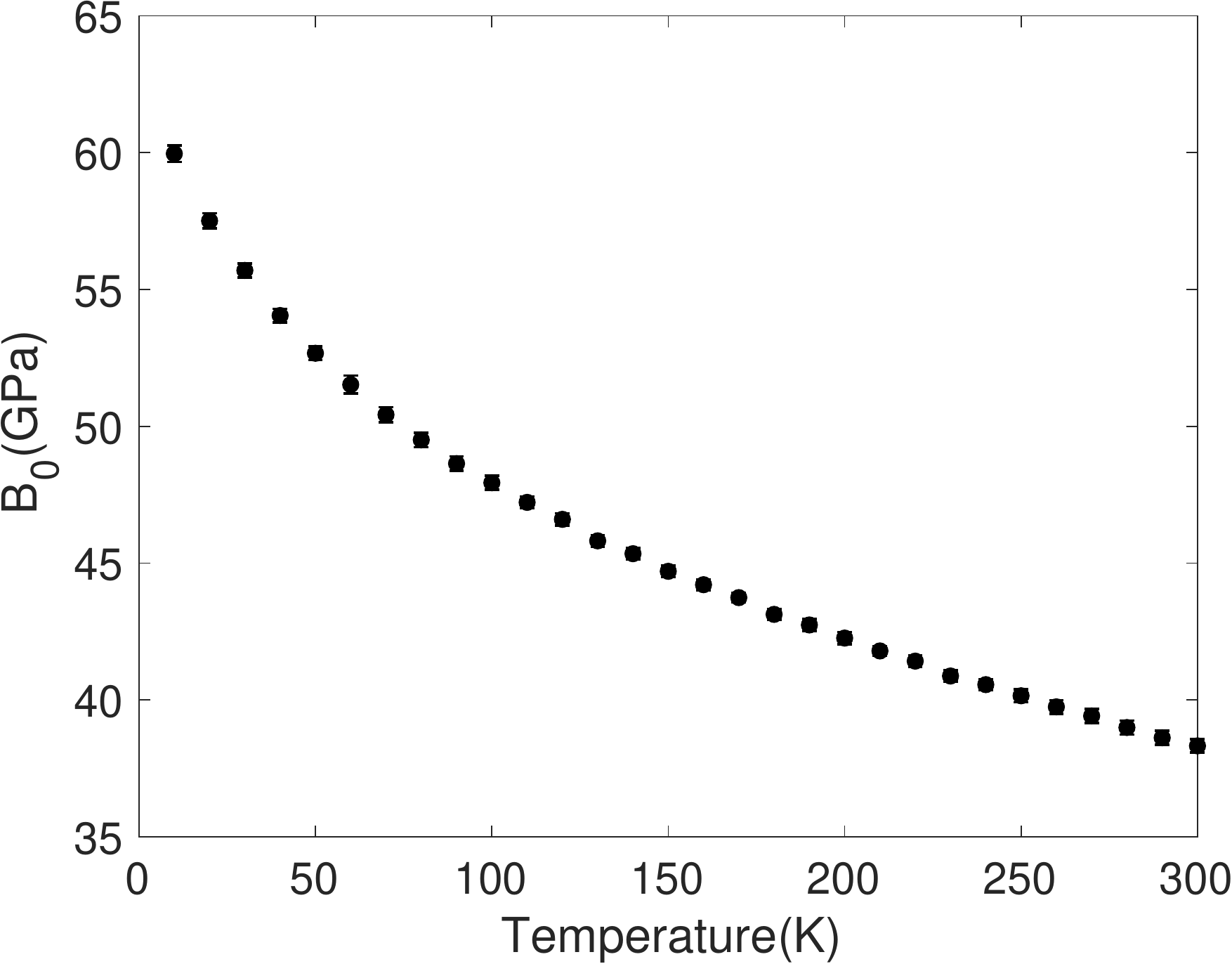}
\subcaption{}
\end{subfigure}
\begin{subfigure}[b]{0.48\textwidth}
\includegraphics[width=0.9\textwidth]{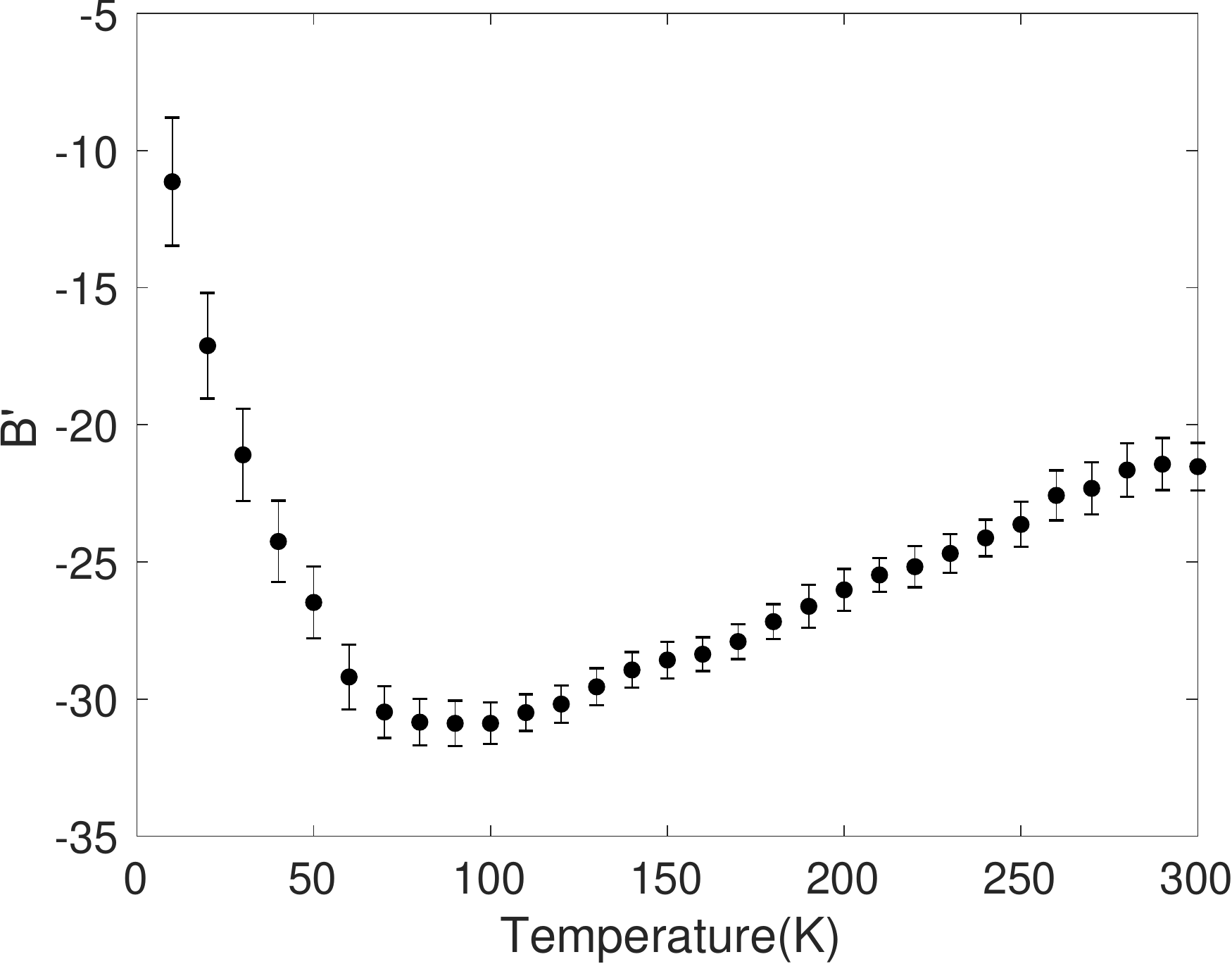}
\subcaption{}
\end{subfigure}
\caption{Temperature dependence of $B_0$ (a) and $B^\prime$ (b) obtained from fitting the third-order Birch-Murnaghan equation of state to crystal volume obtained from molecular dynamics simulation of our idealised model for ScF$_3$. The actual fitting is shown in Figure \ref{fig:PV_MD}} \label{fig:resultsMD}
\end{center}
\end{figure}

We have analysed the pressure dependence of the atomic structure in the simulation over a wide range of temperatures, analysing the results using the standard third-order Birch-Murnaghan equation of state:
\begin{widetext}
\begin{equation}
P(V)=\frac{3B_0}{2}\left[\left(\frac{V_0}{V}\right)^{\frac{7}{3}}-\left(\frac{V_0}{V}\right)^{\frac{5}{3}}\right] \times \left\{1+\frac{3}{4}(B^\prime-4)\left[\left(\frac{V_0}{V}\right)^{\frac{2}{3}}-1\right]\right\}
\label{eq:BMeos}
\end{equation}
\end{widetext}
where $P$ and $V$ are the pressure and volume respectively, and $V_0$ and $B_0$  are the values of $V$ and $B$ at zero pressure. As an expansion of the second-order Birch-Murnaghan equation of state, the third-order equation reduces to the second-order form when $B^\prime = +4$, and many experimental systems actually have values of $B^\prime$ of roughly this size. Data for the functions $V(P)$ showing the fitted equation of state are shown in Figure \ref{fig:PV_MD}, and the fitted results for $B_0$ and $B^\prime$ are shown in Figure \ref{fig:resultsMD}. The temperature-dependence of $B^\prime$ is very similar to that seen in Zn(CN)$_2$ and zeolites \cite{Fang:2013gp, Fang:2013fj, Fang:2014gy}, with a rapid change in the value of $B^\prime$  on heating from zero temperature, until reaching a minimum value (the most negative value) and then a slow change to less negative values on further heating \cite{Fang:2014gy}.  What is astonishing is that our model for ScF$_3$ predicts values of $B^\prime$ as extreme as $-25$. As we noted above, reports of negative values of $B^\prime$ are very rare; such a large negative value is unprecedented.

\begin{figure}[t]
\begin{center}
\begin{subfigure}[b]{0.48\textwidth}
\includegraphics[width=0.9\textwidth]{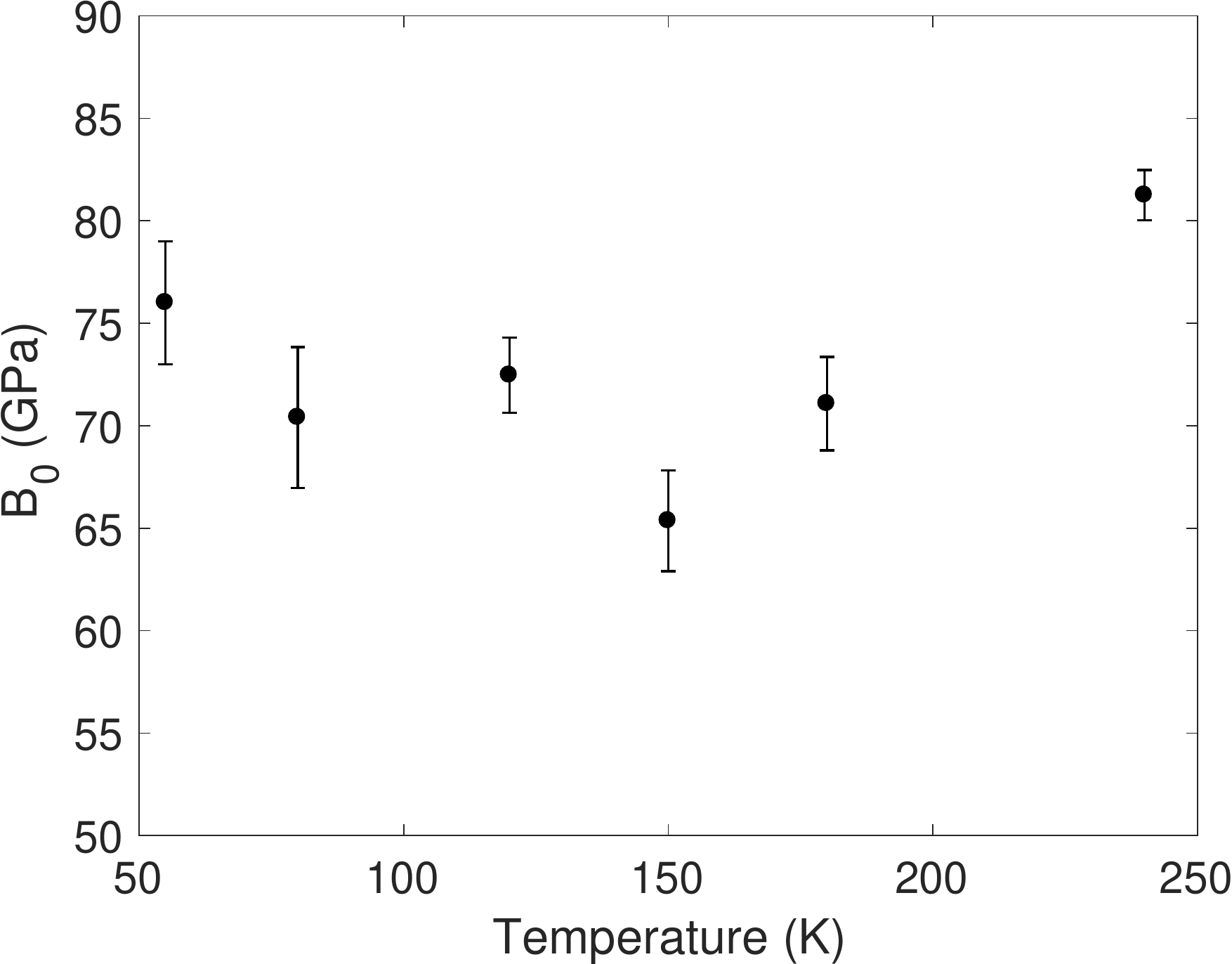}
\subcaption{}
\end{subfigure}
\begin{subfigure}[b]{0.48\textwidth}
\includegraphics[width=0.9\textwidth]{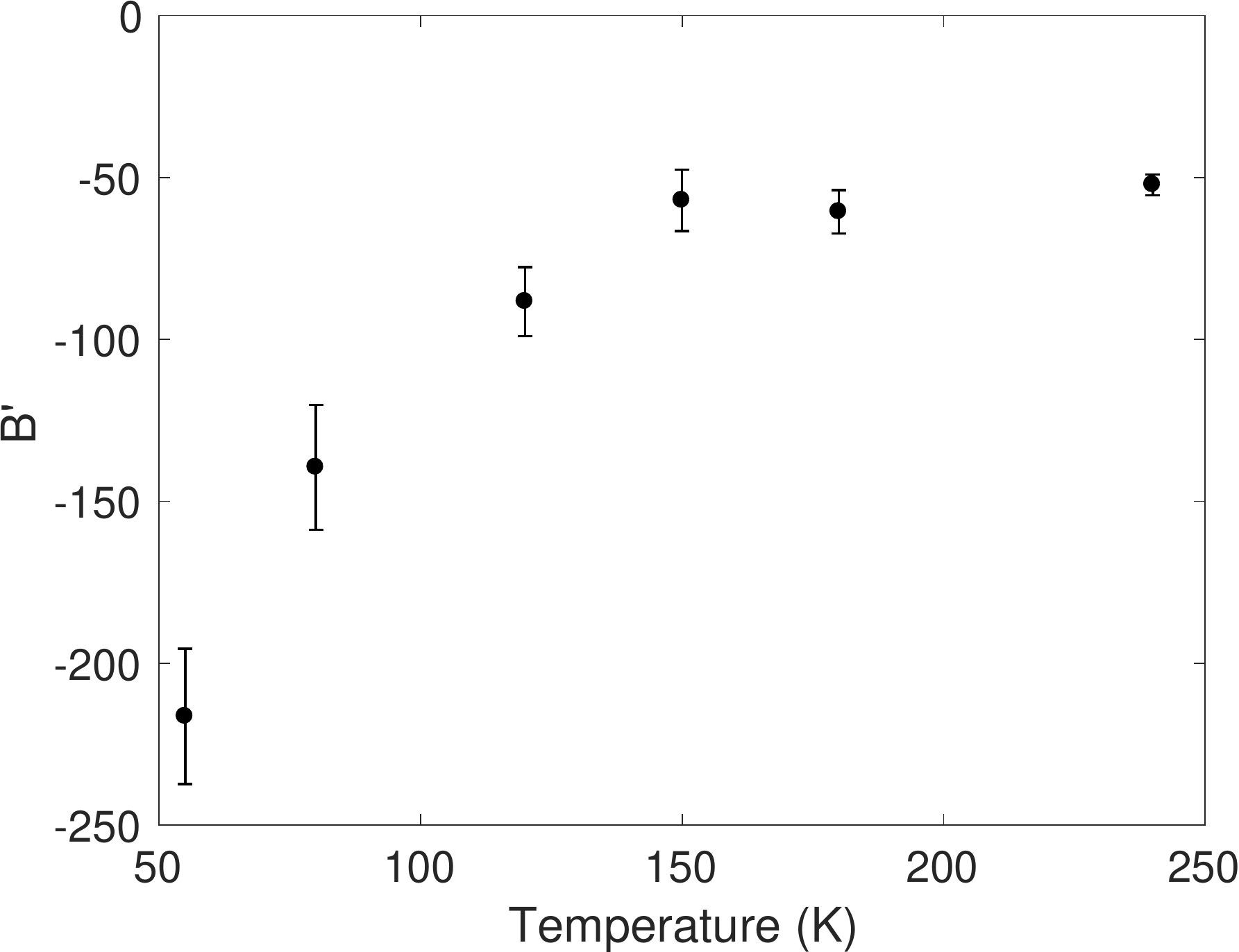}
\subcaption{}
\end{subfigure}
\caption{Temperature dependence of $B_0$ (a) and $B^\prime$ (b) obtained from fitting the third-order Birch-Murnaghan equation of state to experimental data for the crystal volume as obtained from Rietveld analysis of diffraction data. The fitting is shown in Figure \ref{fig:PV_diffraction}.} \label{fig:resultsExp}
\end{center}
\end{figure}

To follow up this prediction we performed neutron powder diffraction measurements of ScF$_3$ under pressure for different temperatures on the GEM diffractometer at the ISIS spallation neutron facility. The diffraction patterns were fitted using the Rietveld method to obtain the lattice parameters and the atomic displacement parameters (all other aspects of the crystal structure are established by symmetry). All details of the experimental and analysis methods are given in the Supplementary Materials, with a representative fit to the data shown in Figure S2.  

The resultant experimental $P(V)$ for different temperatures were fitted by the third-order Birch-Murnaghan equation of state and shown in Figure \ref{fig:PV_diffraction}. The fitted values of $B_0$ and $B^\prime$ are shown in Figure \ref{fig:resultsExp}. The astonishing result is that the extraordinarily-large negative values of $B^\prime$ predicted by simulation are not only reproduced but are \textit{exceeded} in the experimental results. We checked the robustness of this result by trimming the data range used in the fitting, and obtained results each time that are consistent within the estimated standard deviations. It might be argued that the third-order Birch-Murnaghan equation of state was not developed in anticipation of such large negative values of $B^\prime$, given that its derivation involves a Taylor expansion. However, the facts that the function fits the data to within the statistical accuracy, and that the fitting gives consistently large values of $B^\prime$ with relatively small estimated standard errors, are quantitative indications of the remarkable behaviour seen here.

\begin{figure}[t]
\begin{center}
\includegraphics[width=0.45\textwidth]{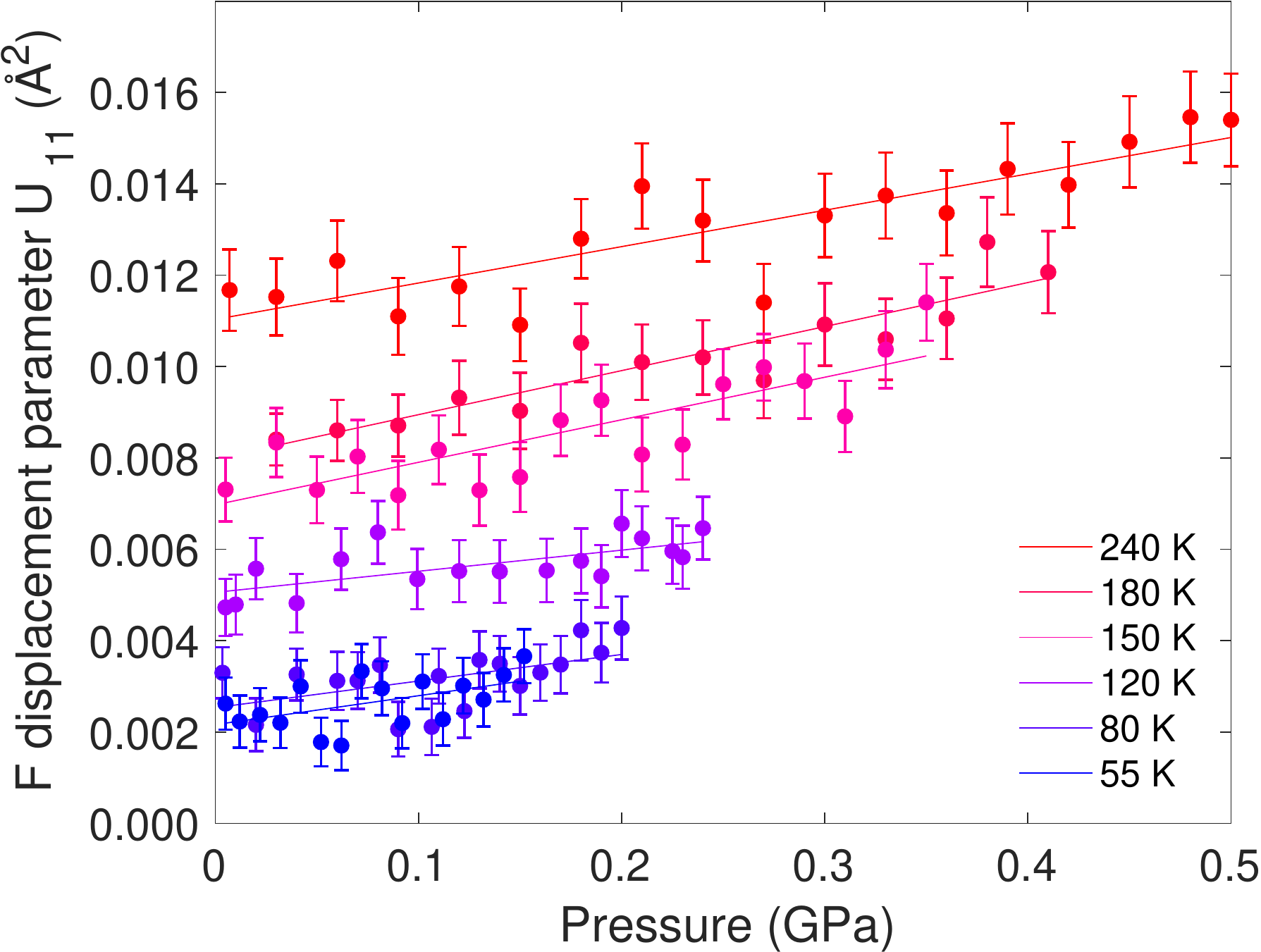}
\caption{Mean-squared transverse displacement of the F atoms obtained from refinement of the crystal structure.}
\label{fig:usq}
\end{center}
\end{figure}

Figure \ref{fig:usq} shows the fitted values of the fluorine transverse atomic displacement parameter -- essentially the mean-square atomic displacement -- as functions of pressure for different temperatures. This corresponds to the large transverse axis of the ellipsoids shown in the crystal structure, Figure \ref{fig:crystal_structure}. At low pressure this quantity increases linearly with temperature as expected if the displacements are associated with thermal motions (the data are presented in Figure S7). Consistent with the ideas presented here, the data presented in Figure \ref{fig:usq} show that the mean-square amplitude of transverse motion of the fluorine atoms \textit{increases} with pressure. On the other hand, in most materials the phonon frequencies increase under pressure due to force constants increasing, and higher frequencies lead to lower mean-square amplitudes of thermal motion. The behaviour shown in Figure \ref{fig:usq} is consistent with the existence of enhanced rotations of the Sc--F bonds on increasing pressure. By contrast the mean-square amplitudes of the other motions (isotropic motion of the Sc atom, longitudinal motion of the F atom) show no pressure dependence (Figure S8). These results are reproduced in the results of the molecular simulations on the simple model (Figure S5).

To explain why the pressure-induced softening of ScF$_3$ is so extreme, we consider our simple potential model. Recall that this has a stiff Sc--F bond, and two terms that control the flexing of the F--Sc--F right angle in the ScF$_6$ octahedron and of the linear Sc--F--Sc angle. Our recent total scattering study of ScF$_6$ \cite{Dove:2019tm} shows that there is considerable flexing of both these angles as compared with corresponding motions in other perovskites. The fact that NTE is largely absent in oxide perovskites suggests that this higher level of flexibility is essential for the mechanism of NTE, and by extension to the enhanced pressure-induced softening seen in ScF$_3$. To confirm this, in separate simulations we increased the parameters governing each angle term to increase the forces opposing bending. We found that increasing either of these parameters leads leads to a clear reduction in NTE, indeed turning the thermal expansion into positive expansivity. The effect is strongest from the F--Sc--F angle term, where increasing the value of the force constant by a factor of only 3.3 is sufficient to drive the thermal expansivity to a positive value\cite{Dove:2019tm}. Our simulations show further that the same changes in the model parameters will reduce the negative size of $B^\prime$. This confirms that the enhanced pressure-induced softening in ScF$_3$ is associated with the high degree of angular flexibility found in this material.

Such large values of $B^\prime$ as seen in this study for ScF$_3$ are extraordinary. Zn(CN)$_2$ only has $B^\prime$  down to values of $-9.0 \pm 0.7$ \cite{Fang:2013fj}. The values of $B^\prime$ seen in this study (Figure \ref{fig:resultsExp}) are so far outside the norm that it seems appropriate to recognise this by calling the effect ``colossal pressure-induced softening''. \footnote{Prof Angus Wilkinson of Georgia Institute of Technology has recently reported to the corresponding author that he has seen similar effects at room temperature in doped samples of ScF$_3$.}

Because of the link we have demonstrated between anomalous thermodynamic properties, the same crystal-engineering techniques used to design negative thermal expansion materials will be useful in discovering additional colossal pressure-induced softening materials. From a technological perspective, pressure-induced softening is particularly relevant to composite materials, where internal stress will be induced by differing thermal expansivities. By showing how extraordinarily large such effects can be, our results emphasise the importance of taking them into account when designing such composites. But we anticipate that colossal pressure-induced softening will also be a powerful tool that can be used to deliberately tune the thermodynamic behaviour and physical properties of these materials.

\section*{Acknowledgments}
GC and LT are grateful for funding from the China Scholarship Council and Queen Mary University of London. Neutron beam time was provided by ISIS under project number RB1820458. Simulations were performed on the HPC Midlands Plus tier-2 system supported by EPSRC (EP/P020232/1) (MTD co-investigator).

\bibliography{scibib}

\begin{thebibliography}{36}%
\makeatletter
\providecommand \@ifxundefined [1]{%
 \@ifx{#1\undefined}
}%
\providecommand \@ifnum [1]{%
 \ifnum #1\expandafter \@firstoftwo
 \else \expandafter \@secondoftwo
 \fi
}%
\providecommand \@ifx [1]{%
 \ifx #1\expandafter \@firstoftwo
 \else \expandafter \@secondoftwo
 \fi
}%
\providecommand \natexlab [1]{#1}%
\providecommand \enquote  [1]{``#1''}%
\providecommand \bibnamefont  [1]{#1}%
\providecommand \bibfnamefont [1]{#1}%
\providecommand \citenamefont [1]{#1}%
\providecommand \href@noop [0]{\@secondoftwo}%
\providecommand \href [0]{\begingroup \@sanitize@url \@href}%
\providecommand \@href[1]{\@@startlink{#1}\@@href}%
\providecommand \@@href[1]{\endgroup#1\@@endlink}%
\providecommand \@sanitize@url [0]{\catcode `\\12\catcode `\$12\catcode
  `\&12\catcode `\#12\catcode `\^12\catcode `\_12\catcode `\%12\relax}%
\providecommand \@@startlink[1]{}%
\providecommand \@@endlink[0]{}%
\providecommand \url  [0]{\begingroup\@sanitize@url \@url }%
\providecommand \@url [1]{\endgroup\@href {#1}{\urlprefix }}%
\providecommand \urlprefix  [0]{URL }%
\providecommand \Eprint [0]{\href }%
\providecommand \doibase [0]{http://dx.doi.org/}%
\providecommand \selectlanguage [0]{\@gobble}%
\providecommand \bibinfo  [0]{\@secondoftwo}%
\providecommand \bibfield  [0]{\@secondoftwo}%
\providecommand \translation [1]{[#1]}%
\providecommand \BibitemOpen [0]{}%
\providecommand \bibitemStop [0]{}%
\providecommand \bibitemNoStop [0]{.\EOS\space}%
\providecommand \EOS [0]{\spacefactor3000\relax}%
\providecommand \BibitemShut  [1]{\csname bibitem#1\endcsname}%
\let\auto@bib@innerbib\@empty
%</preamble>
\bibitem [{\citenamefont {Mary}\ \emph {et~al.}(1996)\citenamefont {Mary},
  \citenamefont {Evans}, \citenamefont {Vogt},\ and\ \citenamefont
  {Sleight}}]{Mary:1996uq}%
  \BibitemOpen
  \bibfield  {author} {\bibinfo {author} {\bibfnamefont {T.~A.}\ \bibnamefont
  {Mary}}, \bibinfo {author} {\bibfnamefont {J.~S.~O.}\ \bibnamefont {Evans}},
  \bibinfo {author} {\bibfnamefont {T.}~\bibnamefont {Vogt}}, \ and\ \bibinfo
  {author} {\bibfnamefont {A.~W.}\ \bibnamefont {Sleight}},\ }\href {\doibase
  10.1126/science.272.5258.90} {\bibfield  {journal} {\bibinfo  {journal}
  {Science}\ }\textbf {\bibinfo {volume} {272}},\ \bibinfo {pages} {90}
  (\bibinfo {year} {1996})}\BibitemShut {NoStop}%
\bibitem [{\citenamefont {Takenaka}(2012)}]{Takenaka:2012cv}%
  \BibitemOpen
  \bibfield  {author} {\bibinfo {author} {\bibfnamefont {K.}~\bibnamefont
  {Takenaka}},\ }\href {\doibase 10.1088/1468-6996/13/1/013001} {\bibfield
  {journal} {\bibinfo  {journal} {Science and Technology of Advanced
  Materials}\ }\textbf {\bibinfo {volume} {13}},\ \bibinfo {pages} {013001}
  (\bibinfo {year} {2012})}\BibitemShut {NoStop}%
\bibitem [{\citenamefont {Romao}\ \emph {et~al.}(2013)\citenamefont {Romao},
  \citenamefont {Miller}, \citenamefont {Whitman}, \citenamefont {White},\ and\
  \citenamefont {Marinkovic}}]{Romao:2013ch}%
  \BibitemOpen
  \bibfield  {author} {\bibinfo {author} {\bibfnamefont {C.~P.}\ \bibnamefont
  {Romao}}, \bibinfo {author} {\bibfnamefont {K.~J.}\ \bibnamefont {Miller}},
  \bibinfo {author} {\bibfnamefont {C.~A.}\ \bibnamefont {Whitman}}, \bibinfo
  {author} {\bibfnamefont {M.~A.}\ \bibnamefont {White}}, \ and\ \bibinfo
  {author} {\bibfnamefont {B.~A.}\ \bibnamefont {Marinkovic}},\ }in\ \href
  {\doibase 10.1016/B978-0-08-097774-4.00425-3} {\emph {\bibinfo {booktitle}
  {Comprehensive Inorganic Chemistry II: From elements to applications}}}\
  (\bibinfo  {publisher} {Elsevier},\ \bibinfo {year} {2013})\ pp.\ \bibinfo
  {pages} {127--151}\BibitemShut {NoStop}%
\bibitem [{\citenamefont {Dove}\ and\ \citenamefont
  {Fang}(2016)}]{Dove:2016bv}%
  \BibitemOpen
  \bibfield  {author} {\bibinfo {author} {\bibfnamefont {M.~T.}\ \bibnamefont
  {Dove}}\ and\ \bibinfo {author} {\bibfnamefont {H.}~\bibnamefont {Fang}},\
  }\href {\doibase 10.1088/0034-4885/79/6/066503} {\bibfield  {journal}
  {\bibinfo  {journal} {Reports on Progress in Physics}\ }\textbf {\bibinfo
  {volume} {79}},\ \bibinfo {pages} {066503} (\bibinfo {year}
  {2016})}\BibitemShut {NoStop}%
\bibitem [{\citenamefont {Goodwin}\ \emph {et~al.}(2008)\citenamefont
  {Goodwin}, \citenamefont {Calleja}, \citenamefont {Conterio}, \citenamefont
  {Dove}, \citenamefont {Evans}, \citenamefont {Keen}, \citenamefont {Peters},\
  and\ \citenamefont {Tucker}}]{Goodwin:2008gb}%
  \BibitemOpen
  \bibfield  {author} {\bibinfo {author} {\bibfnamefont {A.~L.}\ \bibnamefont
  {Goodwin}}, \bibinfo {author} {\bibfnamefont {M.}~\bibnamefont {Calleja}},
  \bibinfo {author} {\bibfnamefont {M.~J.}\ \bibnamefont {Conterio}}, \bibinfo
  {author} {\bibfnamefont {M.~T.}\ \bibnamefont {Dove}}, \bibinfo {author}
  {\bibfnamefont {J.~S.~O.}\ \bibnamefont {Evans}}, \bibinfo {author}
  {\bibfnamefont {D.~A.}\ \bibnamefont {Keen}}, \bibinfo {author}
  {\bibfnamefont {L.}~\bibnamefont {Peters}}, \ and\ \bibinfo {author}
  {\bibfnamefont {M.~G.}\ \bibnamefont {Tucker}},\ }\href {\doibase
  10.1126/science.1151442} {\bibfield  {journal} {\bibinfo  {journal}
  {Science}\ }\textbf {\bibinfo {volume} {319}},\ \bibinfo {pages} {794}
  (\bibinfo {year} {2008})}\BibitemShut {NoStop}%
\bibitem [{\citenamefont {Cairns}\ and\ \citenamefont
  {Goodwin}(2015)}]{Cairns:2015iw}%
  \BibitemOpen
  \bibfield  {author} {\bibinfo {author} {\bibfnamefont {A.~B.}\ \bibnamefont
  {Cairns}}\ and\ \bibinfo {author} {\bibfnamefont {A.~L.}\ \bibnamefont
  {Goodwin}},\ }\href {\doibase 10.1039/C5CP00442J} {\bibfield  {journal}
  {\bibinfo  {journal} {Physical Chemistry Chemical Physics}\ }\textbf
  {\bibinfo {volume} {17}},\ \bibinfo {pages} {20449} (\bibinfo {year}
  {2015})}\BibitemShut {NoStop}%
\bibitem [{\citenamefont {Alderson}\ and\ \citenamefont
  {Alderson}(2007)}]{Alderson:2007bg}%
  \BibitemOpen
  \bibfield  {author} {\bibinfo {author} {\bibfnamefont {A.}~\bibnamefont
  {Alderson}}\ and\ \bibinfo {author} {\bibfnamefont {K.~L.}\ \bibnamefont
  {Alderson}},\ }\href {\doibase 10.1243/09544100JAERO185} {\bibfield
  {journal} {\bibinfo  {journal} {Proceedings of the Institution of Mechanical
  Engineers, Part G: Journal of Aerospace Engineering}\ }\textbf {\bibinfo
  {volume} {221}},\ \bibinfo {pages} {565} (\bibinfo {year}
  {2007})}\BibitemShut {NoStop}%
\bibitem [{\citenamefont {Fang}\ \emph
  {et~al.}(2013{\natexlab{a}})\citenamefont {Fang}, \citenamefont {Dove},
  \citenamefont {Rimmer},\ and\ \citenamefont {Misquitta}}]{Fang:2013ji}%
  \BibitemOpen
  \bibfield  {author} {\bibinfo {author} {\bibfnamefont {H.}~\bibnamefont
  {Fang}}, \bibinfo {author} {\bibfnamefont {M.~T.}\ \bibnamefont {Dove}},
  \bibinfo {author} {\bibfnamefont {L.~H.~N.}\ \bibnamefont {Rimmer}}, \ and\
  \bibinfo {author} {\bibfnamefont {A.~J.}\ \bibnamefont {Misquitta}},\ }\href
  {\doibase 10.1103/PhysRevB.88.104306} {\bibfield  {journal} {\bibinfo
  {journal} {Physical Review B}\ }\textbf {\bibinfo {volume} {88}},\ \bibinfo
  {pages} {104306} (\bibinfo {year} {2013}{\natexlab{a}})}\BibitemShut
  {NoStop}%
\bibitem [{\citenamefont {Fang}\ \emph
  {et~al.}(2013{\natexlab{b}})\citenamefont {Fang}, \citenamefont {Phillips},
  \citenamefont {Dove}, \citenamefont {Tucker},\ and\ \citenamefont
  {Goodwin}}]{Fang:2013fj}%
  \BibitemOpen
  \bibfield  {author} {\bibinfo {author} {\bibfnamefont {H.}~\bibnamefont
  {Fang}}, \bibinfo {author} {\bibfnamefont {A.~E.}\ \bibnamefont {Phillips}},
  \bibinfo {author} {\bibfnamefont {M.~T.}\ \bibnamefont {Dove}}, \bibinfo
  {author} {\bibfnamefont {M.~G.}\ \bibnamefont {Tucker}}, \ and\ \bibinfo
  {author} {\bibfnamefont {A.~L.}\ \bibnamefont {Goodwin}},\ }\href {\doibase
  10.1103/PhysRevB.88.144103} {\bibfield  {journal} {\bibinfo  {journal}
  {Physical Review B}\ }\textbf {\bibinfo {volume} {88}},\ \bibinfo {pages}
  {144103} (\bibinfo {year} {2013}{\natexlab{b}})}\BibitemShut {NoStop}%
\bibitem [{\citenamefont {Fang}\ \emph {et~al.}(2014)\citenamefont {Fang},
  \citenamefont {Dove},\ and\ \citenamefont {Phillips}}]{Fang:2014cp}%
  \BibitemOpen
  \bibfield  {author} {\bibinfo {author} {\bibfnamefont {H.}~\bibnamefont
  {Fang}}, \bibinfo {author} {\bibfnamefont {M.~T.}\ \bibnamefont {Dove}}, \
  and\ \bibinfo {author} {\bibfnamefont {A.~E.}\ \bibnamefont {Phillips}},\
  }\href {\doibase 10.1103/PhysRevB.89.214103} {\bibfield  {journal} {\bibinfo
  {journal} {Physical Review B}\ }\textbf {\bibinfo {volume} {89}},\ \bibinfo
  {pages} {214103} (\bibinfo {year} {2014})}\BibitemShut {NoStop}%
\bibitem [{\citenamefont {Fang}\ and\ \citenamefont
  {Dove}(2013)}]{Fang:2013gp}%
  \BibitemOpen
  \bibfield  {author} {\bibinfo {author} {\bibfnamefont {H.}~\bibnamefont
  {Fang}}\ and\ \bibinfo {author} {\bibfnamefont {M.~T.}\ \bibnamefont
  {Dove}},\ }\href {\doibase 10.1103/PhysRevB.87.214109} {\bibfield  {journal}
  {\bibinfo  {journal} {Physical Review B}\ }\textbf {\bibinfo {volume} {87}},\
  \bibinfo {pages} {214109} (\bibinfo {year} {2013})}\BibitemShut {NoStop}%
\bibitem [{\citenamefont {Tucker}\ \emph {et~al.}(2000)\citenamefont {Tucker},
  \citenamefont {Dove},\ and\ \citenamefont {A}}]{SiObond}%
  \BibitemOpen
  \bibfield  {author} {\bibinfo {author} {\bibfnamefont {M.~G.}\ \bibnamefont
  {Tucker}}, \bibinfo {author} {\bibfnamefont {M.~T.}\ \bibnamefont {Dove}}, \
  and\ \bibinfo {author} {\bibfnamefont {K.~D.}\ \bibnamefont {A}},\
  }\href@noop {} {\bibfield  {journal} {\bibinfo  {journal} {Journal of
  Physics: Condensed Matter}\ }\textbf {\bibinfo {volume} {12}},\ \bibinfo
  {pages} {L425} (\bibinfo {year} {2000})}\BibitemShut {NoStop}%
\bibitem [{\citenamefont {Hui}\ \emph {et~al.}(2005)\citenamefont {Hui},
  \citenamefont {Tucker}, \citenamefont {Dove}, \citenamefont {Wells},\ and\
  \citenamefont {Keen}}]{SrTiO3_RMC}%
  \BibitemOpen
  \bibfield  {author} {\bibinfo {author} {\bibfnamefont {Q.}~\bibnamefont
  {Hui}}, \bibinfo {author} {\bibfnamefont {M.~G.}\ \bibnamefont {Tucker}},
  \bibinfo {author} {\bibfnamefont {M.~T.}\ \bibnamefont {Dove}}, \bibinfo
  {author} {\bibfnamefont {S.~A.}\ \bibnamefont {Wells}}, \ and\ \bibinfo
  {author} {\bibfnamefont {D.~A.}\ \bibnamefont {Keen}},\ }\href@noop {}
  {\bibfield  {journal} {\bibinfo  {journal} {Journal of Physics: Condensed
  Matter}\ }\textbf {\bibinfo {volume} {17}},\ \bibinfo {pages} {S111}
  (\bibinfo {year} {2005})}\BibitemShut {NoStop}%
\bibitem [{\citenamefont {Goodwin}\ \emph {et~al.}(2007)\citenamefont
  {Goodwin}, \citenamefont {Redfern}, \citenamefont {Dove}, \citenamefont
  {Keen},\ and\ \citenamefont {Tucker}}]{SrSnO3_RMC}%
  \BibitemOpen
  \bibfield  {author} {\bibinfo {author} {\bibfnamefont {A.~L.}\ \bibnamefont
  {Goodwin}}, \bibinfo {author} {\bibfnamefont {S.~A.~T.}\ \bibnamefont
  {Redfern}}, \bibinfo {author} {\bibfnamefont {M.~T.}\ \bibnamefont {Dove}},
  \bibinfo {author} {\bibfnamefont {D.~A.}\ \bibnamefont {Keen}}, \ and\
  \bibinfo {author} {\bibfnamefont {M.~G.}\ \bibnamefont {Tucker}},\
  }\href@noop {} {\bibfield  {journal} {\bibinfo  {journal} {Physical Review
  B}\ }\textbf {\bibinfo {volume} {76}},\ \bibinfo {pages} {174114} (\bibinfo
  {year} {2007})}\BibitemShut {NoStop}%
\bibitem [{\citenamefont {Conterio}\ \emph {et~al.}(2008)\citenamefont
  {Conterio}, \citenamefont {Goodwin}, \citenamefont {Tucker}, \citenamefont
  {Keen}, \citenamefont {Dove}, \citenamefont {Peters},\ and\ \citenamefont
  {Evans}}]{Ag3CoCN6_RMC}%
  \BibitemOpen
  \bibfield  {author} {\bibinfo {author} {\bibfnamefont {M.~J.}\ \bibnamefont
  {Conterio}}, \bibinfo {author} {\bibfnamefont {A.~L.}\ \bibnamefont
  {Goodwin}}, \bibinfo {author} {\bibfnamefont {M.~G.}\ \bibnamefont {Tucker}},
  \bibinfo {author} {\bibfnamefont {D.~A.}\ \bibnamefont {Keen}}, \bibinfo
  {author} {\bibfnamefont {M.~T.}\ \bibnamefont {Dove}}, \bibinfo {author}
  {\bibfnamefont {L.}~\bibnamefont {Peters}}, \ and\ \bibinfo {author}
  {\bibfnamefont {J.~S.~O.}\ \bibnamefont {Evans}},\ }\href@noop {} {\bibfield
  {journal} {\bibinfo  {journal} {Journal of Physics: Condensed Matter}\
  }\textbf {\bibinfo {volume} {20}},\ \bibinfo {pages} {255225} (\bibinfo
  {year} {2008})}\BibitemShut {NoStop}%
\bibitem [{\citenamefont {Hu}\ \emph {et~al.}(2016)\citenamefont {Hu},
  \citenamefont {Chen}, \citenamefont {Sanson}, \citenamefont {Wu},
  \citenamefont {Guglieri~Rodriguez}, \citenamefont {Olivi}, \citenamefont
  {Ren}, \citenamefont {Fan}, \citenamefont {Deng},\ and\ \citenamefont
  {Xing}}]{Hu:2016it}%
  \BibitemOpen
  \bibfield  {author} {\bibinfo {author} {\bibfnamefont {L.}~\bibnamefont
  {Hu}}, \bibinfo {author} {\bibfnamefont {J.}~\bibnamefont {Chen}}, \bibinfo
  {author} {\bibfnamefont {A.}~\bibnamefont {Sanson}}, \bibinfo {author}
  {\bibfnamefont {H.}~\bibnamefont {Wu}}, \bibinfo {author} {\bibfnamefont
  {C.}~\bibnamefont {Guglieri~Rodriguez}}, \bibinfo {author} {\bibfnamefont
  {L.}~\bibnamefont {Olivi}}, \bibinfo {author} {\bibfnamefont
  {Y.}~\bibnamefont {Ren}}, \bibinfo {author} {\bibfnamefont {L.}~\bibnamefont
  {Fan}}, \bibinfo {author} {\bibfnamefont {J.}~\bibnamefont {Deng}}, \ and\
  \bibinfo {author} {\bibfnamefont {X.}~\bibnamefont {Xing}},\ }\href {\doibase
  10.1021/jacs.6b02370} {\bibfield  {journal} {\bibinfo  {journal} {Journal of
  the American Chemical Society}\ }\textbf {\bibinfo {volume} {138}},\ \bibinfo
  {pages} {8320} (\bibinfo {year} {2016})}\BibitemShut {NoStop}%
\bibitem [{\citenamefont {Wendt}\ \emph {et~al.}(2019)\citenamefont {Wendt},
  \citenamefont {Bozin}, \citenamefont {Neuefeind}, \citenamefont {Page},
  \citenamefont {Ku}, \citenamefont {Wang}, \citenamefont {Fultz},
  \citenamefont {Tkachenko},\ and\ \citenamefont {Zaliznyak}}]{Wendt:2019it}%
  \BibitemOpen
  \bibfield  {author} {\bibinfo {author} {\bibfnamefont {D.}~\bibnamefont
  {Wendt}}, \bibinfo {author} {\bibfnamefont {E.}~\bibnamefont {Bozin}},
  \bibinfo {author} {\bibfnamefont {J.}~\bibnamefont {Neuefeind}}, \bibinfo
  {author} {\bibfnamefont {K.}~\bibnamefont {Page}}, \bibinfo {author}
  {\bibfnamefont {W.}~\bibnamefont {Ku}}, \bibinfo {author} {\bibfnamefont
  {L.}~\bibnamefont {Wang}}, \bibinfo {author} {\bibfnamefont {B.}~\bibnamefont
  {Fultz}}, \bibinfo {author} {\bibfnamefont {A.~V.}\ \bibnamefont
  {Tkachenko}}, \ and\ \bibinfo {author} {\bibfnamefont {I.~A.}\ \bibnamefont
  {Zaliznyak}},\ }\href@noop {} {\bibfield  {journal} {\bibinfo  {journal}
  {Science Advances}\ }\textbf {\bibinfo {volume} {5}},\ \bibinfo {pages}
  {2748} (\bibinfo {year} {2019})}\BibitemShut {NoStop}%
\bibitem [{\citenamefont {Dove}\ \emph {et~al.}(2019)\citenamefont {Dove},
  \citenamefont {Du}, \citenamefont {Keen}, \citenamefont {Tucker},\ and\
  \citenamefont {Phillips}}]{Dove:2019tm}%
  \BibitemOpen
  \bibfield  {author} {\bibinfo {author} {\bibfnamefont {M.~T.}\ \bibnamefont
  {Dove}}, \bibinfo {author} {\bibfnamefont {J.}~\bibnamefont {Du}}, \bibinfo
  {author} {\bibfnamefont {D.~A.}\ \bibnamefont {Keen}}, \bibinfo {author}
  {\bibfnamefont {M.~G.}\ \bibnamefont {Tucker}}, \ and\ \bibinfo {author}
  {\bibfnamefont {A.~E.}\ \bibnamefont {Phillips}},\ }\href@noop {} {\bibfield
  {journal} {\bibinfo  {journal} {Submitted for publication}\ } (\bibinfo
  {year} {2019})}\BibitemShut {NoStop}%
\bibitem [{\citenamefont {Tsiok}\ \emph {et~al.}(1998)\citenamefont {Tsiok},
  \citenamefont {Brazhkin}, \citenamefont {Lyapin},\ and\ \citenamefont
  {Khvostantsev}}]{Tsiok:1998hf}%
  \BibitemOpen
  \bibfield  {author} {\bibinfo {author} {\bibfnamefont {O.}~\bibnamefont
  {Tsiok}}, \bibinfo {author} {\bibfnamefont {V.}~\bibnamefont {Brazhkin}},
  \bibinfo {author} {\bibfnamefont {A.}~\bibnamefont {Lyapin}}, \ and\ \bibinfo
  {author} {\bibfnamefont {L.}~\bibnamefont {Khvostantsev}},\ }\href {\doibase
  10.1103/PhysRevLett.80.999} {\bibfield  {journal} {\bibinfo  {journal}
  {Physical Review Letters}\ }\textbf {\bibinfo {volume} {80}},\ \bibinfo
  {pages} {999} (\bibinfo {year} {1998})}\BibitemShut {NoStop}%
\bibitem [{\citenamefont {Walker}\ \emph {et~al.}(2007)\citenamefont {Walker},
  \citenamefont {Sullivan}, \citenamefont {Trachenko}, \citenamefont {Bruin},
  \citenamefont {White}, \citenamefont {Dove}, \citenamefont {Tyer},
  \citenamefont {Todorov},\ and\ \citenamefont {Wells}}]{Walker:2007fp}%
  \BibitemOpen
  \bibfield  {author} {\bibinfo {author} {\bibfnamefont {A.~M.}\ \bibnamefont
  {Walker}}, \bibinfo {author} {\bibfnamefont {L.~A.}\ \bibnamefont
  {Sullivan}}, \bibinfo {author} {\bibfnamefont {K.}~\bibnamefont {Trachenko}},
  \bibinfo {author} {\bibfnamefont {R.~P.}\ \bibnamefont {Bruin}}, \bibinfo
  {author} {\bibfnamefont {T.~O.~H.}\ \bibnamefont {White}}, \bibinfo {author}
  {\bibfnamefont {M.~T.}\ \bibnamefont {Dove}}, \bibinfo {author}
  {\bibfnamefont {R.~P.}\ \bibnamefont {Tyer}}, \bibinfo {author}
  {\bibfnamefont {I.~T.}\ \bibnamefont {Todorov}}, \ and\ \bibinfo {author}
  {\bibfnamefont {S.~A.}\ \bibnamefont {Wells}},\ }\href {\doibase
  10.1088/0953-8984/19/27/275210} {\bibfield  {journal} {\bibinfo  {journal}
  {Journal of Physics: Condensed Matter}\ }\textbf {\bibinfo {volume} {19}},\
  \bibinfo {pages} {275210} (\bibinfo {year} {2007})}\BibitemShut {NoStop}%
\bibitem [{\citenamefont {Giddy}\ \emph {et~al.}(1993)\citenamefont {Giddy},
  \citenamefont {Dove}, \citenamefont {Pawley},\ and\ \citenamefont
  {Heine}}]{Giddy:1993ue}%
  \BibitemOpen
  \bibfield  {author} {\bibinfo {author} {\bibfnamefont {A.~P.}\ \bibnamefont
  {Giddy}}, \bibinfo {author} {\bibfnamefont {M.~T.}\ \bibnamefont {Dove}},
  \bibinfo {author} {\bibfnamefont {G.~S.}\ \bibnamefont {Pawley}}, \ and\
  \bibinfo {author} {\bibfnamefont {V.}~\bibnamefont {Heine}},\ }\href
  {\doibase 10.1107/S0108767393002545} {\bibfield  {journal} {\bibinfo
  {journal} {Acta Crystallographica Section A Foundations of Crystallography}\
  }\textbf {\bibinfo {volume} {49}},\ \bibinfo {pages} {697} (\bibinfo {year}
  {1993})}\BibitemShut {NoStop}%
\bibitem [{\citenamefont {Hammonds}\ \emph {et~al.}(1996)\citenamefont
  {Hammonds}, \citenamefont {Dove}, \citenamefont {Giddy}, \citenamefont
  {Heine},\ and\ \citenamefont {Winkler}}]{Hammonds:1996wy}%
  \BibitemOpen
  \bibfield  {author} {\bibinfo {author} {\bibfnamefont {K.~D.}\ \bibnamefont
  {Hammonds}}, \bibinfo {author} {\bibfnamefont {M.~T.}\ \bibnamefont {Dove}},
  \bibinfo {author} {\bibfnamefont {A.~P.}\ \bibnamefont {Giddy}}, \bibinfo
  {author} {\bibfnamefont {V.}~\bibnamefont {Heine}}, \ and\ \bibinfo {author}
  {\bibfnamefont {B.}~\bibnamefont {Winkler}},\ }\href {\doibase
  10.2138/am-1996-9-1003} {\bibfield  {journal} {\bibinfo  {journal} {American
  Mineralogist}\ }\textbf {\bibinfo {volume} {81}},\ \bibinfo {pages} {1057}
  (\bibinfo {year} {1996})}\BibitemShut {NoStop}%
\bibitem [{\citenamefont {Pryde}\ and\ \citenamefont
  {Dove}(1998)}]{Tridymite_RUMs}%
  \BibitemOpen
  \bibfield  {author} {\bibinfo {author} {\bibfnamefont {A.~K.~A.}\
  \bibnamefont {Pryde}}\ and\ \bibinfo {author} {\bibfnamefont {M.~T.}\
  \bibnamefont {Dove}},\ }\href@noop {} {\bibfield  {journal} {\bibinfo
  {journal} {Physics and Chemistry of Minerals}\ }\textbf {\bibinfo {volume}
  {26}},\ \bibinfo {pages} {171} (\bibinfo {year} {1998})}\BibitemShut
  {NoStop}%
\bibitem [{\citenamefont {Heine}\ \emph {et~al.}(1999)\citenamefont {Heine},
  \citenamefont {Welche},\ and\ \citenamefont {Dove}}]{Heine:1999vk}%
  \BibitemOpen
  \bibfield  {author} {\bibinfo {author} {\bibfnamefont {V.}~\bibnamefont
  {Heine}}, \bibinfo {author} {\bibfnamefont {P.~R.~L.}\ \bibnamefont
  {Welche}}, \ and\ \bibinfo {author} {\bibfnamefont {M.~T.}\ \bibnamefont
  {Dove}},\ }\href@noop {} {\bibfield  {journal} {\bibinfo  {journal} {Journal
  of the American Ceramic Society}\ }\textbf {\bibinfo {volume} {82}},\
  \bibinfo {pages} {1793} (\bibinfo {year} {1999})}\BibitemShut {NoStop}%
\bibitem [{\citenamefont {Drymiotis}\ \emph {et~al.}(2004)\citenamefont
  {Drymiotis}, \citenamefont {Ledbetter}, \citenamefont {Betts}, \citenamefont
  {Kimura}, \citenamefont {Lashley}, \citenamefont {Migliori}, \citenamefont
  {Ramirez}, \citenamefont {Kowach},\ and\ \citenamefont
  {Van~Duijn}}]{Drymiotis:2004kc}%
  \BibitemOpen
  \bibfield  {author} {\bibinfo {author} {\bibfnamefont {F.}~\bibnamefont
  {Drymiotis}}, \bibinfo {author} {\bibfnamefont {H.}~\bibnamefont
  {Ledbetter}}, \bibinfo {author} {\bibfnamefont {J.}~\bibnamefont {Betts}},
  \bibinfo {author} {\bibfnamefont {T.}~\bibnamefont {Kimura}}, \bibinfo
  {author} {\bibfnamefont {J.}~\bibnamefont {Lashley}}, \bibinfo {author}
  {\bibfnamefont {A.}~\bibnamefont {Migliori}}, \bibinfo {author}
  {\bibfnamefont {A.}~\bibnamefont {Ramirez}}, \bibinfo {author} {\bibfnamefont
  {G.}~\bibnamefont {Kowach}}, \ and\ \bibinfo {author} {\bibfnamefont
  {J.}~\bibnamefont {Van~Duijn}},\ }\href {\doibase
  10.1103/PhysRevLett.93.025502} {\bibfield  {journal} {\bibinfo  {journal}
  {Physical Review Letters}\ }\textbf {\bibinfo {volume} {93}},\ \bibinfo
  {pages} {025502} (\bibinfo {year} {2004})}\BibitemShut {NoStop}%
\bibitem [{\citenamefont {Chapman}\ and\ \citenamefont
  {Chupas}(2007)}]{Chapman:2007fg}%
  \BibitemOpen
  \bibfield  {author} {\bibinfo {author} {\bibfnamefont {K.~W.}\ \bibnamefont
  {Chapman}}\ and\ \bibinfo {author} {\bibfnamefont {P.~J.}\ \bibnamefont
  {Chupas}},\ }\href {\doibase 10.1021/ja073791e} {\bibfield  {journal}
  {\bibinfo  {journal} {Journal of the American Chemical Society}\ }\textbf
  {\bibinfo {volume} {129}},\ \bibinfo {pages} {10090} (\bibinfo {year}
  {2007})}\BibitemShut {NoStop}%
\bibitem [{\citenamefont {Greve}\ \emph {et~al.}(2010)\citenamefont {Greve},
  \citenamefont {Martin}, \citenamefont {Lee}, \citenamefont {Chupas},
  \citenamefont {Chapman},\ and\ \citenamefont {Wilkinson}}]{Greve:2010bu}%
  \BibitemOpen
  \bibfield  {author} {\bibinfo {author} {\bibfnamefont {B.~K.}\ \bibnamefont
  {Greve}}, \bibinfo {author} {\bibfnamefont {K.~L.}\ \bibnamefont {Martin}},
  \bibinfo {author} {\bibfnamefont {P.~L.}\ \bibnamefont {Lee}}, \bibinfo
  {author} {\bibfnamefont {P.~J.}\ \bibnamefont {Chupas}}, \bibinfo {author}
  {\bibfnamefont {K.~W.}\ \bibnamefont {Chapman}}, \ and\ \bibinfo {author}
  {\bibfnamefont {A.~P.}\ \bibnamefont {Wilkinson}},\ }\href {\doibase
  10.1021/ja106711v} {\bibfield  {journal} {\bibinfo  {journal} {Journal of the
  American Chemical Society}\ }\textbf {\bibinfo {volume} {132}},\ \bibinfo
  {pages} {15496} (\bibinfo {year} {2010})}\BibitemShut {NoStop}%
\bibitem [{\citenamefont {Li}\ \emph {et~al.}(2011)\citenamefont {Li},
  \citenamefont {Tang}, \citenamefont {Mu{\~n}oz}, \citenamefont {Keith},
  \citenamefont {Tracy}, \citenamefont {Abernathy},\ and\ \citenamefont
  {Fultz}}]{Li:2011dn}%
  \BibitemOpen
  \bibfield  {author} {\bibinfo {author} {\bibfnamefont {C.~W.}\ \bibnamefont
  {Li}}, \bibinfo {author} {\bibfnamefont {X.}~\bibnamefont {Tang}}, \bibinfo
  {author} {\bibfnamefont {J.~A.}\ \bibnamefont {Mu{\~n}oz}}, \bibinfo {author}
  {\bibfnamefont {J.~B.}\ \bibnamefont {Keith}}, \bibinfo {author}
  {\bibfnamefont {S.~J.}\ \bibnamefont {Tracy}}, \bibinfo {author}
  {\bibfnamefont {D.~L.}\ \bibnamefont {Abernathy}}, \ and\ \bibinfo {author}
  {\bibfnamefont {B.}~\bibnamefont {Fultz}},\ }\href {\doibase
  10.1103/PhysRevLett.107.195504} {\bibfield  {journal} {\bibinfo  {journal}
  {Physical Review Letters}\ }\textbf {\bibinfo {volume} {107}},\ \bibinfo
  {pages} {195504} (\bibinfo {year} {2011})}\BibitemShut {NoStop}%
\bibitem [{\citenamefont {Morelock}\ \emph {et~al.}(2013)\citenamefont
  {Morelock}, \citenamefont {Greve}, \citenamefont {Gallington}, \citenamefont
  {Chapman},\ and\ \citenamefont {Wilkinson}}]{Morelock:2013gi}%
  \BibitemOpen
  \bibfield  {author} {\bibinfo {author} {\bibfnamefont {C.~R.}\ \bibnamefont
  {Morelock}}, \bibinfo {author} {\bibfnamefont {B.~K.}\ \bibnamefont {Greve}},
  \bibinfo {author} {\bibfnamefont {L.~C.}\ \bibnamefont {Gallington}},
  \bibinfo {author} {\bibfnamefont {K.~W.}\ \bibnamefont {Chapman}}, \ and\
  \bibinfo {author} {\bibfnamefont {A.~P.}\ \bibnamefont {Wilkinson}},\ }\href
  {\doibase 10.1063/1.4836855} {\bibfield  {journal} {\bibinfo  {journal}
  {Journal of Applied Physics}\ }\textbf {\bibinfo {volume} {114}},\ \bibinfo
  {pages} {213501} (\bibinfo {year} {2013})}\BibitemShut {NoStop}%
\bibitem [{\citenamefont {Handunkanda}\ \emph {et~al.}(2015)\citenamefont
  {Handunkanda}, \citenamefont {Curry}, \citenamefont {Voronov}, \citenamefont
  {Said}, \citenamefont {Guzm{\'a}n-Verri}, \citenamefont {Brierley},
  \citenamefont {Littlewood},\ and\ \citenamefont
  {Hancock}}]{Handunkanda:2015dc}%
  \BibitemOpen
  \bibfield  {author} {\bibinfo {author} {\bibfnamefont {S.~U.}\ \bibnamefont
  {Handunkanda}}, \bibinfo {author} {\bibfnamefont {E.~B.}\ \bibnamefont
  {Curry}}, \bibinfo {author} {\bibfnamefont {V.}~\bibnamefont {Voronov}},
  \bibinfo {author} {\bibfnamefont {A.~H.}\ \bibnamefont {Said}}, \bibinfo
  {author} {\bibfnamefont {G.~G.}\ \bibnamefont {Guzm{\'a}n-Verri}}, \bibinfo
  {author} {\bibfnamefont {R.~T.}\ \bibnamefont {Brierley}}, \bibinfo {author}
  {\bibfnamefont {P.~B.}\ \bibnamefont {Littlewood}}, \ and\ \bibinfo {author}
  {\bibfnamefont {J.~N.}\ \bibnamefont {Hancock}},\ }\href {\doibase
  10.1103/PhysRevB.92.134101} {\bibfield  {journal} {\bibinfo  {journal}
  {Physical Review B}\ }\textbf {\bibinfo {volume} {92}},\ \bibinfo {pages}
  {134101} (\bibinfo {year} {2015})}\BibitemShut {NoStop}%
\bibitem [{\citenamefont {Wang}\ \emph
  {et~al.}(2015{\natexlab{a}})\citenamefont {Wang}, \citenamefont {Wang},
  \citenamefont {Sun}, \citenamefont {Deng}, \citenamefont {Shi}, \citenamefont
  {Lu}, \citenamefont {Hu},\ and\ \citenamefont {Zhang}}]{Wang:2015gm}%
  \BibitemOpen
  \bibfield  {author} {\bibinfo {author} {\bibfnamefont {L.}~\bibnamefont
  {Wang}}, \bibinfo {author} {\bibfnamefont {C.}~\bibnamefont {Wang}}, \bibinfo
  {author} {\bibfnamefont {Y.}~\bibnamefont {Sun}}, \bibinfo {author}
  {\bibfnamefont {S.}~\bibnamefont {Deng}}, \bibinfo {author} {\bibfnamefont
  {K.}~\bibnamefont {Shi}}, \bibinfo {author} {\bibfnamefont {H.}~\bibnamefont
  {Lu}}, \bibinfo {author} {\bibfnamefont {P.}~\bibnamefont {Hu}}, \ and\
  \bibinfo {author} {\bibfnamefont {X.}~\bibnamefont {Zhang}},\ }\href
  {\doibase 10.1111/jace.13676} {\bibfield  {journal} {\bibinfo  {journal}
  {Journal of the American Ceramic Society}\ }\textbf {\bibinfo {volume}
  {98}},\ \bibinfo {pages} {2852} (\bibinfo {year}
  {2015}{\natexlab{a}})}\BibitemShut {NoStop}%
\bibitem [{\citenamefont {Wang}\ \emph
  {et~al.}(2015{\natexlab{b}})\citenamefont {Wang}, \citenamefont {Wang},
  \citenamefont {Sun}, \citenamefont {Shi}, \citenamefont {Deng}, \citenamefont
  {Lu}, \citenamefont {Hu},\ and\ \citenamefont {Zhang}}]{Wang:2015eo}%
  \BibitemOpen
  \bibfield  {author} {\bibinfo {author} {\bibfnamefont {L.}~\bibnamefont
  {Wang}}, \bibinfo {author} {\bibfnamefont {C.}~\bibnamefont {Wang}}, \bibinfo
  {author} {\bibfnamefont {Y.}~\bibnamefont {Sun}}, \bibinfo {author}
  {\bibfnamefont {K.}~\bibnamefont {Shi}}, \bibinfo {author} {\bibfnamefont
  {S.}~\bibnamefont {Deng}}, \bibinfo {author} {\bibfnamefont {H.}~\bibnamefont
  {Lu}}, \bibinfo {author} {\bibfnamefont {P.}~\bibnamefont {Hu}}, \ and\
  \bibinfo {author} {\bibfnamefont {X.}~\bibnamefont {Zhang}},\ }\href
  {\doibase 10.1016/j.jmat.2015.02.001} {\bibfield  {journal} {\bibinfo
  {journal} {Journal of Materiomics}\ }\textbf {\bibinfo {volume} {1}},\
  \bibinfo {pages} {106} (\bibinfo {year} {2015}{\natexlab{b}})}\BibitemShut
  {NoStop}%
\bibitem [{\citenamefont {Oba}\ \emph {et~al.}(2019)\citenamefont {Oba},
  \citenamefont {Tadano}, \citenamefont {Akashi},\ and\ \citenamefont
  {Tsuneyuki}}]{Oba:2019bi}%
  \BibitemOpen
  \bibfield  {author} {\bibinfo {author} {\bibfnamefont {Y.}~\bibnamefont
  {Oba}}, \bibinfo {author} {\bibfnamefont {T.}~\bibnamefont {Tadano}},
  \bibinfo {author} {\bibfnamefont {R.}~\bibnamefont {Akashi}}, \ and\ \bibinfo
  {author} {\bibfnamefont {S.}~\bibnamefont {Tsuneyuki}},\ }\href {\doibase
  10.1103/PhysRevMaterials.3.033601} {\bibfield  {journal} {\bibinfo  {journal}
  {Physical Review Materials}\ ,\ \bibinfo {pages} {1}} (\bibinfo {year}
  {2019})}\BibitemShut {NoStop}%
\bibitem [{Note1()}]{Note1}%
  \BibitemOpen
  \bibinfo {note} {The other being the simple rocksalt structure.}\BibitemShut
  {Stop}%
\bibitem [{\citenamefont {Fang}\ and\ \citenamefont
  {Dove}(2014)}]{Fang:2014gy}%
  \BibitemOpen
  \bibfield  {author} {\bibinfo {author} {\bibfnamefont {H.}~\bibnamefont
  {Fang}}\ and\ \bibinfo {author} {\bibfnamefont {M.~T.}\ \bibnamefont
  {Dove}},\ }\href {\doibase 10.1088/0953-8984/26/11/115402} {\bibfield
  {journal} {\bibinfo  {journal} {Journal of Physics: Condensed Matter}\
  }\textbf {\bibinfo {volume} {26}},\ \bibinfo {pages} {115402} (\bibinfo
  {year} {2014})}\BibitemShut {NoStop}%
\bibitem [{Note2()}]{Note2}%
  \BibitemOpen
  \bibinfo {note} {Prof Angus Wilkinson of Georgia Institute of Technology has
  recently reported to the corresponding author that he has seen similar
  effects at room temperature in doped samples of ScF$_3$.}\BibitemShut {Stop}%
\end{thebibliography}%

\end{document}